# The Role of Digital Payments in Driving Regional Economic Growth: *A Panel Data Analysis with Structural Break*[*][1]

Wishnu Badrawani[a], Citra Amanda[a], Novi Maryaningsih[b], Carla Sheila Wulandari[a]

[a]Bank Indonesia Institute, Bank Indonesia, Indonesia
[b]Payment System Policy Department, Bank Indonesia, Indonesia

**Abstract**

Using a panel payment system dataset of thirty-three provinces in Indonesia, we examine the impact of digital payment on the regional economy, considering structural breaks induced by unprecedented events and policies. Digital payments were determined to significantly affect regional income and consumption before and after the identified breakpoint, with the impact greater following the break. Employing a novel method for structural break analysis within interactive effects panel data, we demonstrate that the break in retail payment models is due to COVID-19, and the break in the wholesale payment model is associated with the central bank's payment system policy.

---

[1] The views expressed in this article are those of the author(s) and do not necessarily represent the views or policies of Bank Indonesia. Corresponding Author: Wishnu Badrawani, e-mail: w_badrawani@bi.go.id.

# 1. INTRODUCTION

Digitalization has transformed almost all aspects of life, including economics and culture; however, it can be a double-edged sword. Digitalization, because of innovation, particularly in financial and payment systems, has not only expanded accessibility and inclusivity for individuals and businesses to the financial systems but also created new opportunities that drive economic growth along with its potential drawbacks (Dedola et al., 2023). Innovation in digital payments can be seen in various payment instruments or platforms used in the economy, electronically and digitally, such as debit and credit payment, money transfers using electronic devices or platforms, electronic and digital money, quick response (QR) codes, and any payment method that avoid the use of cash.

According to East Ventures (2024), the value of the digital economy in Indonesia in 2030 is projected to reach US$ 360 billion; however, Indonesia's digital economy only shares 9% of the total GDP, which is the lowest among Southeast Asia (SEA)-6 (Tech for Good Institute, 2023). To accelerate the digital economy's contribution, Indonesia requires technology innovation and infrastructure to be broadened in digital finance and payments. The blueprint for Indonesia's digital economy and the national payment system blueprint (BSPI) has led to rapid innovations in digital payment technology, supported by the government-led expansion of digital infrastructure, which has garnered positive responses from businesses and the public. Bank Indonesia's data shows that digital transaction volume (retail card payment and electronic money) in December 2023 reached 2,311 million transactions, an average increase of 10 % in the last three years. According to the World Bank (2022), digital payment users have significantly increased since the COVID-19 pandemic. The changed behavior of individuals in using digital payment was confirmed to be influenced by both the COVID-19 pandemic and policies promoted by the government and the central bank (Badrawani, 2022).

The adoption of digital financial services in payment systems plays an important role in the economy (Lubis et al., 2019). This condition can be seen in several economic sectors in Indonesia. For example, in the trade sector, the accelerated adoption of digital payments triggered by the COVID-19 pandemic has contributed positively by facilitating more efficient transactions and increasing consumer accessibility to goods and services. Adopting digital payment services is also becoming an increasingly important trend in the transport and warehousing sector. The increasingly common use of prepaid cards or payment platforms across various modes of transport speeds up the passenger loading-unloading process and reduces reliance on cash. This phenomenon is also happening more intensely in the logistics sector, especially in retail logistics. The widespread adoption of digital payments, particularly during the COVID-19 pandemic, in these two sectors has had a significant positive impact in improving business process efficiency and convenience for businesses and consumers, as well as in the accommodation and food beverages sector and other services, confirming the findings of Javaid et al. (2024) who found that the digitalization enhances the culture of the industry.

However, despite these broad sectoral trends, the extent to which digital payments drive economic activity varies significantly across regions due to differences in economic structure, digital infrastructure, and policy implementation. For instance, provinces with higher levels of digital penetration and a more developed service sector may benefit more from digital payment adoption than those with lower levels of digital infrastructure or economic diversification. Past studies have predominantly examined the effects of digital payments on economic growth at the national level (Christianti, 2024; Wasiaturrahma & Kurniasari, 2021), potentially overlooking these regional disparities. To address this gap, our study focuses on exploring the role of digital payment in enhancing regional economic growth using provincial-level panel data across 33 provinces in Indonesia from 2019 to 2023. Our results find that digital innovation contributes positively to driving regional economic growth. Our further examination suggests that the positive impact of digital innovation also affects household consumption. We also explore potential structural breaks in the model that may be triggered by the COVID-19 pandemic and central bank policy related to the payment system, which may have influenced digital payment adoption and economic outcomes at the regional level.

This research is expected to contribute to the literature on payment systems and regional economics in two ways. Firstly, this research focuses on studying the role of digital payment innovation on regional economic growth at the provincial level. Secondly, this research will also examine the possibility of a structural break that may be triggered by central bank policy and other unprecedented events. For policymakers, this research is also expected to provide feedback for strengthening payment system policies, especially regarding its role in the central bank's policy mix framework (Warjiyo & Juhro, 2022). The remainder of this paper is organized as follows. Section 2 presents the literature review and theoretical framework. Section 3 describes the research methodology. Results and discussions in Section 4. Section 5 presents the conclusions and policy recommendations.

## 2. LITERATURE REVIEW

Digital payment innovation is essential for accelerating economic development, transforming financial ecosystems, and enhancing economic activities. Wong et al. (2020) found that economic development in OECD countries is significantly influenced by cashless payments, particularly debit card payments. The effect of digital payment on economic growth is transmitted through three main channels: consumption, government, and capital. According to Dubey & Purnanandam (2023), the effect of digital payment was more pronounced in financially underdeveloped countries than in developed countries. This finding contrasts with Pang et al. (2022), who discovered that the effect was more pronounced in developed countries, with e-money being the only significant instrument. According to Ravikumar et al. (2019), digital payments have a substantial short-term effect on economic development but do not have a long-term impact. As per Aguilar et al. (2024), digital payment adoption is positively correlated with the expansion of GDP per capita and the reduction of informal sector employment. According to Rooj & Sengupta (2020), payment platforms significantly impact the economy, both income and price levels, not only retail payments but also wholesale payments RTGS, with a bidirectional causality.

Digital payments significantly enhance economic growth by broadening access to banking services and improving financial inclusion (Shpanel-Yukhta, 2022). Evidence indicates that digital innovation significantly impacts economic growth through financial inclusion, particularly by enhancing financial ecosystems (Daud & Ahmad, 2023), suggesting that consolidation initiatives should prioritize enhancing financial ecosystems by expanding digital infrastructure. Liu et al. (2021) confirmed the relationship between digital payments and increased financial inclusion, which significantly fosters economic growth by facilitating access to credit that supports the performance of small and medium-sized enterprises (SMEs). This finding corroborates Varlamova et al. (2020), who found that digitalization affects all elements of the economy and finance, altering the behavior of economic players. The availability of the internet and the widespread use of mobile phones are found to influence saving behavior; hence, having a bank account is associated with a higher probability of saving. The positive influence of digital payment on economic growth through financial inclusion was more prevalent in emerging countries (Erlando et al., 2020; Van et al., 2021). The impact of digital payment innovation on financial inclusion indicates that it may serve as an alternative solution to reduce poverty rates and income inequality, especially in developing countries (Inoue, 2024; Liang & Li, 2023). In addition, digital payments contribute to economic development by fostering innovation, reducing fraud, and enhancing efficiency in the global payment system (Aldaas, 2021). The increase in sustainable economic development and consumption is also significantly influenced by digital payment (Zhou, 2022). Many studies indicate that digital payment innovation significantly influences government financial performance; for example, Mouna & Jarboui (2022) in their study in the Middle Eastern and North African countries (MENA) countries.

Research conducted by Apeti & Edoh (2023) on a large sample of 104 developing countries reveals that mobile money contributes to increased government income from direct and indirect tax revenue, supported by Kumar (2024). Gorshkov (2022) and Raouf (2022) discovered that nations adopting digital payment instruments in government sector transactions had experienced enhanced government revenue, especially among middle-upper-class income groups. On the government expenditure side, digitalization of the government budget also helps to increase budget disbursement governance during program implementation, such as in the case of Australia's social support program to reduce gambling (Greenacre et al., 2023) and the governance of MSMEs (Bhatnagar et al., 2024). Additionally, the research conducted by Setor et al. (2021) across 111 developing countries indicates that digital payments contribute to enhanced transparency and reduced corruption. This finding is supported by Kubatko et al. (2023) and Dobrovolska & Rozhkova (2024), despite some drawbacks associated with mobile payment that may increase corruption in the short term in developing countries with civil-law jurisdictions, as noted by Tang et al. (2022). The digitization of payments in the government sector boosts income and transparency and facilitates interoperability across various payment and digital platforms with e-commerce, expanding the digitalization of trade in the area and enabling cross-border transactions (Dedola et al., 2023; Deloitte, 2024). Consequently, the country's sustainable development and natural resource utilization are enhanced because of the country's increased governance (Castro & Lopes, 2022).

The study by Wong et al. (2020) indicates that in OECD countries, the adoption of digital payments empirically enhances economic growth by facilitating business expansion and increasing economies of scale. Patra & Sethi (2024) corroborate these findings, indicating that the innovation of retail e-payment enhances transactions and increases efficiency, which subsequently stimulates economic growth, consumption, and trade. The rise of digital payment systems has markedly accelerated as a result of the COVID-19 pandemic, contributing positively to economic growth (Saroy et al., 2022). The adoption of digital payments has been shown to increase market potential for businesses and strengthen the digital economy ecosystem (Liang & Li, 2023). This finding supports the assertion by Raharja et al. (2020) that digital payment innovation improves business opportunities through the provision of digital payment solutions and diverse financial services, thereby facilitating access to a wider market. Aguilar et al. (2024) conducted a study across 101 countries, encompassing both advanced economies and emerging market and developing economies (EMDEs), indicating that the adoption of digital payments significantly influences economic growth, with a more pronounced effect in emerging economies. Specifically, a one percentage point increase in digital payment usage is associated with a 0.05 percentage point increase in per capita GDP growth. Digital financial innovation enhances the availability of financial services and lowers the expenses related to digital infrastructure, including bank branches, ATMs, and POS terminals, thereby fostering economic growth (Chakraborty & Gogia, 2022; Misati et al., 2024). The influence of digital payments within the digital economy on economic growth is significant, as they enhance productivity, facilitate innovations that create new opportunities, such as monitoring sustainability progress, and optimize resource utilization (Javaid et al., 2024).

In Indonesia, research conducted by Mashabi & Wasiaturrahma (2021) indicates that card payments (ATM and debit card, credit card and electronic money transactions) have a favorable long-term effect on economic growth, a finding supported by Christianti (2024) except for credit card transactions. However, this contradicts Ravikumar et al. (2019), who found that digital payments had a significant short-term impact. According to Farida et al. (2023), the COVID-19 pandemic accelerated the adoption of electronic payments, positively influencing Indonesia's economic growth. Digital payments as a key digital economy component have fostered an upsurge in e-commerce transactions and promoted business expansion (Tanjung et al., 2022). Figure 1 provides an overview of three potential pathways through which digital payments impact regional economies based on the existing literature.

---------------------------

Insert Figure 1

---------------------------

This study develops three main transmission channels extended from Wong et al. (2020). First, the saving channel enhances financial inclusion, enabling households and SMEs to increase savings and improve financial resilience. Second, the investment channel lowers operational costs and expands market access for merchants, fostering business expansion and economic productivity. Third, the government expenditure channel facilitates tax collection efficiency, allowing for increased public revenue that supports pro-growth policies. Collectively, these channels accelerate economic growth by boosting income and consumption, reinforcing the role of digital payment systems in transforming financial ecosystems.

## 3. DATA AND METHODOLOGY
### 3.1 Data

This research employs quarterly data of Bank Indonesia's payment system statistics datasets from 2019 to 2023, examining panel data across 33 provinces. We also utilize macroeconomic data from Indonesia's Economic and Finance Statistics (SEKI) and the Central Bureau of Statistics for control variables. Digital payment is defined as any non-cash payment instrument, including RTGS, debit or credit cards, electronic or digital money, mobile payments, and internet payments, utilized for transactions from an account or for purchasing goods and services online (Aguilar et al., 2024). This study categorizes digital payment transactions into two types: retail payments and wholesale payments. Retail payment transactions encompass ATM/debit cards, credit cards, and electronic money (card-based and server-based e-money, commonly called e-wallet or mobile wallet), transacted through various payment channels and platforms. RTGS transactions represent wholesale payments. Our analysis focuses on the effects of digital payment innovation through metrics of the nominal value of digital payment transactions.

Policies and unprecedented events that may cause structural breaks related to the relationship between digital payment innovations and economic growth during the observed period may be identified as follows:

- Implementation of the Quick Response Code Indonesian Standard (QRIS), launched in late 2019. This policy seeks to enhance interoperability and efficiency by enabling consumers to perform contactless payment transactions and transfer funds to users of various payment operators, including server-based e-money and mobile banking.
- The COVID-19 pandemic was officially declared to spread in Indonesia in March 2020. It has demonstrated significant impacts on various aspects of individuals' lives, including social and economic dimensions, as well as on businesses and government institutions.
- Bank Indonesia Regulation (PBI) No. 22/23/PBI/2020, dated December 29, 2020, pertains to Payment Systems and came into effect in July 2021. This policy consolidates and streamlines various regulations on payment system industries by integrating the licensing process and payment system provisions under a single legal framework.
- BIFAST (Bank Indonesia Fast Payment) was launched in December 2021 as a real-time settlement system for retail payments.

*3.2   Empirical Model*

This study examines the impact of digital payment innovation on regional economic growth through a panel data analysis of 33 provinces in Indonesia, utilizing the following empirical model:

$$y_{it} = a_0 + b_1 Digital_{it} + b_2 X_{it} + e_t \qquad (1)$$

where $y_t$ is the regional real Gross Domestic Product (GDP) and consumption. This study uses the real GDP of provinces to measure the impact of payment system innovation on regional economic growth, as suggested by many in the existing literature (Growiec, 2023; Wong et al., 2020).

*Digital* refers to the vector of non-cash payment platforms (overall card payments, ATM/debit card, credit card, e-money, and RTGS) in the value of transactions. $X_t$ is a vector of control variables that includes inflation, government spending, capital, and human capital. Including the control variables allows the model to capture the impact of macroeconomy aggregate, particularly monetary, fiscal, investment, and human capital. Those control variables have been widely used in economic research to examine the determinants of a country's growth rate (Aguilar et al., 2024; Lau & Yip, 2019; Wong et al., 2020).

Additional variables have been incorporated into the model to prevent the bias caused by omitted variables. According to Yusgiantoro et al. (2018) and Broby (2021), digital innovation substantially impacts cost efficiency, resulting in the closure of unnecessary bank offices and infrastructures. Consequently, this study proxies the impact of technological innovation and financial infrastructure by examining the interaction between payment terminals (EDCs or POS machines) and retailers adopting digital payments, following the digital economic growth literature (Rath & Hermawan, 2019; Tran & Wang, 2023).

*3.3   Methodology*

This study employs a panel data model that can distinguish the time-varying effect of digital payment adoption from provincial characteristics, which impact the regional economy simultaneously, following Ditzen et al. (2024) and Karavias et al. (2023). The econometric model provides a new toolbox to examine the existence of (unknown) structural breaks in panel data models with interactive effects. With the presence of estimated breaks, we rearrange the equation (1) in the following way:

$$y_{it} = a_0 + b_1(t \leq k)X_{it} + b_2 I(t > k)X_{it} + b_3 \lambda_{it} + u_{it} \qquad (2)$$

where the dependent variable $y_{it}$ is the GDP of province $i$ at time $t$. The independent variables are collected in $X_{it}$, a vector that contains the digital payments as the variable of interest with structural breaks, and $\lambda_{it}$, control variables of the analysis without structural breaks. The digital payments variable being observed in the analysis consists of overall card payment transactions, ATM/debit card, credit card, e-money, and RTGS, and control variables consisting of inflation, government spending, capital, and human capital.

The above econometric approach provides several benefits to answer the research questions. First, the model can capture the time-varying effect of how each explanatory variable may affect differently in a particular period. The slope coefficient of each explanatory variable is captured in $\beta_1$ before the break date $k$ and $\beta_2$ after the break so that the difference between $\beta_2$ and $\beta_1$ captures the net change in coefficients due to the structural break. The parameters of interest, $\beta_1$, $\beta_2$, and $k$, determined from the data, will be shown below. Secondly, this method can also capture the unobserved province-specific variation that may impact regional economic growth. This heterogeneity is indicated by differences in the types of socioeconomic and demographic components supporting the model, including inflation rates, government expenditure and investment rates, and human capital.

Next, the robustness checks have been conducted to ensure the validity of the empirical results. This study adds other related explanatory variables into the model that represent digital infrastructure and ecosystems in each province to avoid omitted variable bias. Interaction variables are utilized to represent the three channels: saving, government, and investment, as illustrated in Figure 1. Notably, this practice is common across a wide variety of disciplines in economics and finance to mitigate omitted variable and endogeneity issues in the model (Aguilar et al., 2024; Wong et al., 2020).

## 4. RESULT

### 4.1 Pra-estimation

#### 4.1.1 Descriptive statistics

Table 1 indicates the descriptive statistics of the balanced panel dataset from equation (2), which consists of 660 observations. The table summarizes the statistics for the regional economy and the explanatory variables in our study, including mean, standard deviation, minimum, maximum, and cross-sectional correlation refer to Pesaran (2021) and unit root test for each variable. All variables used have been stationary, using the p-stationary test of P. Chen et al. (2022) and Karavias (2023) in panel data analysis.

The mean credit card volume is negative, indicating that the use of credit cards for transactions has been reduced during the observation. ATM/Debit cards were the only retail payments higher than the RTGS, and both payments were the top payment platforms used by individuals and businesses.

---------------------------
Insert Table 1
---------------------------

#### 4.1.2 Break Testing and Estimation

Initially, we ascertain whether there are any breaks during the observation period. Considering the COVID-19 pandemic and the existence of payment system-related policies, we do not wish to infer that all of them resulted in breaks, following Ditzen et al. (2024). Instead, we regard the quantity of breaks as unknown.

---------------------------
Insert Table 2
---------------------------

Table 2 displays estimated breakpoints for each break and the 95 percent confidence intervals. The breaks are carefully estimated, with extremely narrow confidence intervals encompassing only a few months before and after. The anticipated breakpoints for models of regional GDP and consumption are similar, occurring in mid-2020. The estimated break date corresponds to a major related event and is immediately important. The suggested break may be the result of the pandemic. President Jokowi officially declared the COVID-19 pandemic in Indonesia on 2 March 2020, in response to the coronavirus outbreak. Cases were identified before March 2020, with 6,575 COVID-19 patients recorded by 19 April 2020. This crisis was followed by the implementation of the Large-Scale Social Restrictions (PSBB) policy, which included limitations on cash usage and an increase in non-cash transactions (Semeru, 2020).

The structural break may also result from the central bank's implementation of QRIS, a platform enabling consumers to make payments via various digital payment providers (e-money and mobile banking), thereby promoting non-cash payment methods. The models, however, suggested that most of the break date for retail payments will occur in the first quarter of 2020, with a 95% confidence level. Additionally, considering that QRIS was implemented in August 2019, we concur that the breakpoint is linked to the COVID-19 pandemic, as evidenced by previous studies (Badrawani, 2024; Karavias, 2022; Kiyota, 2021). The break date in retail payments reflects the changing behavior of consumers and merchants in conducting transactions and the increased growth in non-cash transactions for safety reasons.

The recommended breakpoint for wholesale payment (RTGS) is the first quarter of 2022, which is significantly distant from the COVID-19 outbreak. The implementation of BIFAST in December 2021 may explain this finding, as the maximum cap for BIFAST transactions coincides with the minimum limit for RTGS transactions.[2] Consequently, RTGS must compete in the market alongside the newly established retail real-time settlement platform, BIFAST. The effect of central bank policy enactment that causes structural breaks aligns with Mhd Ruslan & Mokhtar (2020) and Badrawani (2025).

---------------------------
Insert Figure 2
---------------------------

Figure 2 illustrates the retail and wholesale payment instruments, the anticipated structural break, and the regional consumption to facilitate comprehension of the findings presented in Table 2. Regional consumption and retail digital payment, particularly ATM/debit card and e-money transactions, are positively correlated, as evidenced by the rapid decline that was followed by a gradual increase in consumption. The estimated breakpoint occurs between 2020Q1 and 2020Q3. In contrast, wholesale payment (RTGS) and regional consumption are shown to reach a breakpoint between 2021Q4 and 2022Q2.

---

[2] The minimum transaction threshold for RTGS is Rp100 million, whereas the maximum transaction limit for BIFAST is Rp250 million.

*4.1.3  Main Panel Regression Result*

The baseline results of the panel regression estimate for the regional economy as dependent variables (DV) are presented in Tables 3 and 4, as previously mentioned in Eq. (1). The explanatory variables' breakpoint, which indicates the potential for structural breaks due to policy changes or external shocks, is associated with real GDP and consumption and is subsequently analyzed across various alternative models.

We begin with the results in Table 3 for panel data of all 33 provinces, with the regional GDP as the dependent variable. The regional GDP is significantly and positively associated with the usage of digital payments. We observe that the regional economy (GDP) is positively impacted by all retail payment instruments, including ATM/debit cards, credit cards, e-money, and overall card payments (refer to Table 3, columns (1) to (4)). Notably, we observe that the impact of retail digital payments on the regional GDP was statistically significant, with a positive coefficient before the break (second quarter of 2020). Subsequently, the influence of retail digital payments on the regional GDP increased after the suggested break.

Conversely, the regional GDP is not impacted by wholesale payment instruments (RTGS) that commonly facilitate the transfer of money for business purposes or investment, as illustrated in column (5) of Table 3. The RTGS platform has no substantial impact on regional GDP, possibly due to the recently implemented real-time transfer platform by Bank Indonesia, BIFAST. This finding is corroborated by explanations from focus group discussions with government officials and banking representatives, indicating that many government transactions were previously conducted via RTGS but have shifted to BIFAST and that some commercial transactions are being shifted from RTGS to BIFAST as well[3]. However, the RTGS platform significantly impacts regional real GDP when analyzed through transaction volume or frequency[4].

---------------------------
Insert Table 3
---------------------------

---------------------------
Insert Table 4
---------------------------

---

[3] The RTGS platform has considerably impacted regional real GDP when examined over an extended timeframe (2005 to 2023) using a time series analysis method (available upon request). Due to data availability, the analysis of the extended data using panel data is limited.

[4] The panel regression results of trannsaction volume are not reported here to conserve space, but they are available upon request.

Table 4 replicates the panel regression exercise for regional consumption. Higher regional consumption is associated with greater digital payment penetration, as illustrated in Table 4. According to columns (6) to (10), retail and wholesale payments positively impact regional consumption. We discovered that the impact of ATM/debit card transactions and overall card transactions on regional consumption was statistically significant and positive before the break. Subsequently, the impact increased even further after the break date, which was suggested to be within the second quarter of 2020. Initially, the e-money transaction did not significantly impact regional consumption; however, it became significantly positive after the break. The regional consumption was positively impacted by wholesale payments before and after the suggested breakpoint in the first quarter of 2022. The suggested breakpoint may be associated with the BIFAST introduction in December 2021. The overall card payment instruments, ATM/debit cards, and RTGS exhibited a higher impact on consumption than on GDP. This finding may indicate that retail payment instruments are more effective in promoting regional consumption than GDP growth. The positive impact of digital payment uptake on economic development supports the findings of Rooj & Sengupta (2020) and Aguilar et al. (2024). The finding of observed behavioral shifts post-break in our study aligns with findings in other regions, including European nations (Wisniewski et al., 2024) and Asian countries (Mai et al., 2024; Mhd Ruslan & Mokhtar, 2020).

## 4.2 Robustness Check

### 4.2.1 Digital Payment and Saving Channel

Our conceptual framework, as mentioned in Figure 1, suggests that one potential channel in which digital payment can influence the regional economy is through savings, which are boosted by financial inclusion. We measure savings using third-party funds that consist of current accounts, savings accounts, and term deposit accounts, which then interact with digital payment transactions. Table 5 illustrates that the regional GDP is significantly positively influenced by the interaction of digital payments and savings, particularly for ATM/debit cards, credit cards, and all card payments, before and after the break.

Table 6 shows the panel regression exercise for regional consumption replicated from Table 5. Higher regional consumption is associated with greater savings and digital payments, as illustrated in Table 6. According to columns (6) to (10), most retail payments, except e-money, positively impact regional GDP when interacting with savings. The impact of the interaction of digital payments and savings on consumption was found to be significantly positive, and the impact strengthened after the break. Several factors can explain the positive association between the interaction of digital payments and savings and the regional economy. Digital innovation in payment services can promote the adoption of digital payment instruments, thereby enhancing financial inclusion and increasing aggregate savings. According to Beck (2020), digital innovation in payment services can promote digital payment usage, thereby enhancing financial literacy and inclusion. Integrating it with other digital platforms in the ecosystem (e.g., e-commerce) attracts a greater number of individuals to engage with the digital financial system and augment overall savings (Yusgiantoro et al., 2020). Second, financial inclusion is not only supported by fintech companies but also by the emerging digital banks (i.e., Sea Bank and Jenius) and the newly digital approach of conventional banks, for example, through digital banking platforms such as MyBCA and Livin Mandiri, customers no longer need to visit branch offices to access banking services. Moreover, a massive and global digital financial ecosystem can be accessed via a bank's SuperApp application, which includes front-to-end services, making it easier for users to manage all their financial needs Chen et al. (2024). Lastly, digital payment innovation plays a role in expanding financial inclusion in traditional sectors. For example, a regional development bank in East Java province integrates the laku pandai[5] business model with auto-debt features, especially in traditional markets and school canteens, which can help accelerate the adoption of digital finance in communities previously underserved by the modern banking system.

-------------------------
Insert Table 5
-------------------------

-------------------------
Insert Table 6
-------------------------

The insignificant influence of e-money in the saving channel could be attributed to the fact that, unlike ATM/debit cards, the funds in e-money are generated from a personal savings account. Thus, the aggregate value of savings accounts is reduced, with only a small portion of e-money shared in the total payment system. Furthermore, the insignificant impact of wholesale payment on regional GDP, both directly and via the saving channel, is evidenced by the infrequent use of RTGS for daily transactions. On the other hand, the interaction between RTGS and savings was found to have a positive and significant effect on regional consumption. The reason for this is unclear; however, it is suggested that the cash transfer of social safety net programs in the region was allocated to provincial and municipal authorities via RTGS transfer, subsequently deposited into the savings accounts of designated household recipients within the banking system and ultimately used by the households. For more detail on social assistance programs' cash transfers during and after COVID-19, see Setyawan (2023) and Anwar et al. (2024), notwithstanding several issues that arose in the program implementation.

### 4.2.2 Digital Payment and Government Channel

Another potential channel through which digital payment can affect the regional economy is through government channels. To study this channel, we measure government revenue using its spending, assuming the pattern of government income follows its expenditure. Table 7 shows that retail payments positively influence the regional economy through government channels before and after the break, consistent with previous studies (Apeti & Edoh, 2023; Mouna & Jarboui, 2022). Table 8 shows the panel regression exercise for regional consumption replicating Table 7. According to columns (6) to (10), all retail payments that interact with the regional budget positively impact regional consumption. The impact of the interaction of digital payments and the government's effect on consumption was also significantly positive, with its impact increasing after the break.

---

[5] *Laku Pandai* is a non-branch financial service aimed at promoting financial inclusion, offering banking and other financial services via partnerships with agents, facilitated by information technology resources.

      Several reasons can explain this positive relationship. First, payment of taxes and budget disbursement using digital platforms is proven to support more efficient regional financial management with a faster and more governed process (Raouf, 2022; Setor et al., 2021). For example, the significant role of e-money in influencing regional economy through government channels is confirmed by focus group discussions with the Acceleration Project of Regional Digitalization (TP2DD) officials in East Java and North Sumatera Provinces that are held on 17 July 2024 and 4 October 2024, indicating that many merchants and other economic agents utilize online platforms, such as e-wallets, to pay taxes, particularly the ultra-micro businesses. Second, the focus group discussion confirms that nearly all regional tax and fee payments are conducted electronically. A small proportion (35.0%) is handled via semi-digital channels, while 63.48% is digitally processed. The role of digital payment in facilitating tax collection is consistent with Aguilar et al. (2024); hence, streamlining the transaction process aids in the collection of more revenue, such as in the toll road gate (Mashabi & Wasiaturrahma, 2021), which subsequently promotes economic development. Lastly, switching the payment method from a manual to a digital platform could save the government IDR 53.5 billion per year in operational costs (lesson learned from East Java province). The government initiative aimed at optimizing regional government expenditure through government credit cards may explain the notable impact of credit cards on regional GDP and consumption within the government sector, following Wasiaturrahma & Kurniasari (2021). The interaction between RTGS and government positively influences regional consumption, whereas it does not impact regional GDP. This condition may be linked to money transfers via RTGS in government accounts related to expenditures encouraging regional consumption, such as cash transfers for nationwide social safety net programs.

---------------------------
Insert Table 7
---------------------------
---------------------------
Insert Table 8
---------------------------

*4.2.3 Digital Payment and Capital Channel*

The interaction variable between digital payments and capital is a further possible channel for evaluating the influence of digital payment innovation on the regional economy. Table 9 illustrates that the regional GDP is significantly positively influenced by the interaction between digital payments and capital formation before and after the break. The indicated payment platforms include ATM/debit cards, credit cards, e-money, and all card payments, except for RTGS (see Table 9, columns (1) to (5)). Notably, this study supports Wong et al. (2020), who discovered a positive impact of the interaction between digital payment and capital on economic development. This finding can be explained by multiple factors: First, digital payment platforms enable access to the digital economy ecosystem, including e-commerce and ride-hailing services, via banking or fintech digital payment solutions. For instance, e-commerce platforms like Tokopedia and ride-hailing services like Gojek accept various e-money and mobile banking providers. The digital innovation of commercial banks facilitates customers' purchasing of various goods and services across different e-commerce and digital platforms through their digital banking systems, thereby providing wider access to the digital market for MSMEs. Second, adopting digital payments helps merchants effectively manage cash flow and gain access to additional financial services.

The mobile banking platforms and e-wallets offer a range of financial products, including loans, insurance, and Letters of Credit (L/C) for export-import arrangements. The platforms provide various payment channels supporting ultra micro (UMI)[6] merchants in their daily operations, including on-us and off-us transaction options. In the current digital landscape, UMI merchants who do not have bank accounts due to the bank's stringent regulations can access financial systems via fintech companies. This fintech platform facilitates access to digital ecosystems that offer various services, including e-commerce, insurance, and bulk payment services, thereby providing advantages supporting business expansion. Furthermore, integrating digital payment platforms with innovations in financial services, such as the Laku Pandai scheme, can help UMI merchants broaden their business and attract a larger customer base.

-------------------------

Insert Table 9

-------------------------

-------------------------

Insert Table 10

-------------------------

---

[6] Ultra Micro (UMI) is a subset of micro, small, and medium enterprises (MSMEs) with limited resources, informal operations, and low capitalization (<250 million revenue per year) (Affandi et al., 2024).

Table 10 displays the panel regression results of regional consumption and the interaction between digital payments and capital, replicated from Table 9. Higher regional consumption is associated with greater capital and digital payment. According to columns (6) to (10), all retail payments and RTGS that interact with regional capital positively impact regional consumption, with the influence of credit cards being higher on the regional GDP rather than consumption in the capital channel. The impact of the interaction of digital payments and capital on consumption was found to be significantly positive, with its impact becoming stronger after the break.

The influence of e-money was also higher on the regional GDP than consumption in the capital channel. We do not have a precise explanation for this condition; however, we suggest this condition may be related to the transaction using these retail payments associated with the transfer of funds that are related to the spending of capital goods such as tools, electricity, land/building materials, or transport equipment (i.e., motorcycles), that eventually support regional economic growth than consumption. We found that the interaction of RTGS and capital formation positively affects regional consumption while it does not affect the regional GDP. Despite no solid evidence to explain this condition, money transfers using RTGS in the capital channel are found to be associated with transfers that promote local consumption rather than economic growth (Martín Belmonte et al., 2021).

*4.2.4 Digital Payment and Digital Infrastructure*

Theoretically, the baseline results may suffer from various econometric issues, such as omitted variable bias and outliers. Consequently, additional robustness assessments are implemented to guarantee the validity and consistency of the findings. To validate our baseline results, we include additional digital infrastructure variables into the model: the number of electronic data captures (EDC) or point-of-sale (POS) terminals and the interaction of EDC and merchants as a proxy of the digital economy ecosystem. This approach allows the model to avoid the omitted variable bias. Notably, the positive relationship between regional economy and digital payment is consistent throughout all the models, as shown in Tables 11 and 12. This finding further confirms that the baseline results are not biased due to the missing of certain variables compared to those included. In addition, the estimated result is found to be consistent before and after the break.

In addition to payment instruments, digital innovation is also represented by digital infrastructures and ecosystems. Table 11 illustrates that the regional GDP is significantly influenced by digital infrastructure and ecosystem measures: EDC, merchants, and the ratio of EDC per merchant. The coefficient of the digital infrastructure variable, as reported in columns (1) to (3) of Table 11, indicates that the regional economy is considerably impacted by digital infrastructure within economic agents, representing a digital economy ecosystem.

This result further confirms that the regional economy is considerably impacted by digitalization, which enhances the economic agents by adding payment infrastructure advancement in their outlets, transforming the traditional economy into a digital economy ecosystem, following Daud & Ahmad (2023) and Misati et al. (2024). The findings of the regional economic equation, incorporating multiple digital infrastructures, aligned with the baseline estimate before and after the break. Panel regression results show that a unit increase in card payments has a positive and significant influence on the regional economy with the inclusion of the ratio of EDC per merchant variable as of 2.13 percent, with the size increasing after the break date as 2.66 percent.

-------------------------
Insert Table 11
-------------------------
-------------------------
Insert Table 12
-------------------------

Higher regional consumption is also associated with digital payments infrastructure and economic agents, as illustrated in Table 12. According to columns (4) to (6), the digital infrastructure and ecosystem positively affect regional consumption. The estimated results of regional consumption equations incorporating diverse digital infrastructures are consistent with the main panel result (Table 4) before and after the suggested break. To ensure the robustness check of the panel regression result, we also employ a panel regression with random effect, indicating that the estimated result is consistent with the main result[7].

## 5. Conclusion and Policy Recomendation

This study offers novel evidence regarding the influence of digital payments on Indonesia's regional economy from 2019 to 2023. Our empirical findings show a positive and significant relationship between digital payment platforms, both retail and wholesale, and regional GDP and consumption. Second, the impact of digital payments on regional GDP and consumption was statistically significant and positive before the suggested break related to the pandemic and central bank policy; subsequently, this impact increased even further after the break. Third, we confirm the influence of digital payments on regional economics transmitted via three primary channels: saving, government, and capital. Specifically, all retail payment instruments positively and significantly impact regional GDP and consumption via these channels, except for e-money in the saving channel, which has no substantial effect. The large payment platforms positively impact regional consumption but do not influence regional GDP in those three channels. Digital infrastructure availability and the economic agent's adoption of digital payments significantly influence regional economics.

Our contribution provides cross-province evidence regarding the relationship between digital payments and regional economics, which can be further examined. As a policy recommendation, authorities should prioritize the adoption of retail payment instruments that can optimize the saving, government, and capital channels, as well as broaden the availability of digital infrastructure and merchants' adoption of digital payments to expand the digital economy ecosystem. In this manner, the country can optimize the potential benefits of the digital economy ecosystem's development to stimulate regional economic growth.

---

[7] The panel regression results of random effect are not reported here to conserve space, but they are available upon request.

      We acknowledge that our findings are not entirely free of limitations. Digital payment adoption may be more prevalent in more developed provinces, with a concentration in Java Island and a small portion of the western and eastern regions. Consequently, the results may not be generalizable to all provinces due to the varying economic structures and levels of technology adoption. The provincial panel data of payment transactions was derived from the locality of the financial account proprietor, which does not necessarily coincide with the location of each transaction. Furthermore, it is recommended that more detailed and granular data be used for each province examination.

Martín Belmonte, S., Puig, J., Roca, M., & Segura, M. (2021). Crisis Mitigation through Cash Assistance to Increase Local Consumption Levels—A Case Study of a Bimonetary System in Barcelona, Spain. *Journal of Risk and Financial Management*, *14*(9), 430. https://doi.org/10.3390/JRFM14090430/S1

Mashabi, M., & Wasiaturrahma, W. (2021). Analysis of the effect of electronic-based payment systems and economic growth in Indonesia. *Jurnal Ilmu Ekonomi Terapan)*, *6*(1), 97–121. https://doi.org/10.20473/jiet.v6i1.26287

Mhd Ruslan, S. M., & Mokhtar, K. (2020). Structural break and consumer prices: the case of Malaysia. *Cogent Business and Management*, *7*(1). https://doi.org/10.1080/23311975.2020.1767328

Misati, R., Osoro, J., Odongo, M., & Abdul, F. (2024). Does digital financial innovation enhance financial deepening and growth in Kenya? *International Journal of Emerging Markets*, *19*(3). https://doi.org/10.1108/IJOEM-09-2021-1389

Mouna, A., & Jarboui, A. (2022). Understanding the link between government cashless policy, digital financial services and socio-demographic characteristics in the MENA countries. *International Journal of Sociology and Social Policy*, *42*(5–6). https://doi.org/10.1108/IJSSP-12-2020-0544

Pang, Y.-X., Ng, S.-H., & Lau, W.-T. (2022). Digital Cashless Payments and Economic Growth: Evidence from CPMI Countries. *Capital Markets Review*, *30*(2).

Patra, B., & Sethi, N. (2024). Does digital payment induce economic growth in emerging economies? The mediating role of institutional quality, consumption expenditure, and bank credit. *Information Technology for Development*, *30*(1). https://doi.org/10.1080/02681102.2023.2244465

Pesaran, M. H. (2021). General diagnostic tests for cross-sectional dependence in panels. *Empirical Economics*, *60*(1), 13–50. https://doi.org/10.1007/S00181-020-01875-7/METRICS

Raharja, S. J., Sutarjo, S., Muhyi, H. A., & Herawaty, T. (2020). Digital Payment as an Enabler for Business Opportunities: A Go-Pay Case Study. *Review of Integrative Business and Economics Research*, *9*(1). http://buscompress.com/journal-home.html

Raouf, E. (2022). The impact of financial inclusion on tax revenue in EMEA countries: A threshold regression approach. *Borsa Istanbul Review*, *22*(6), 1158–1164. https://doi.org/10.1016/J.BIR.2022.08.003

Rath, B. N., & Hermawan, D. (2019). Do information and communication technologies foster economic growth in Indonesia? *Buletin Ekonomi Moneter Dan Perbankan*, *22*(1). https://doi.org/10.21098/bemp.v22i1.1041

Ravikumar, T., Suresha, B., Sriram, M., & Rajesh, R. (2019). Impact of digital payments on economic growth: Evidence from India. *International Journal of Innovative Technology and Exploring Engineering*, *8*(12). https://doi.org/10.35940/ijitee.L3432.1081219
21

# Tables and Figures

Figure 1 Transmission channel of digital payment on the regional economy

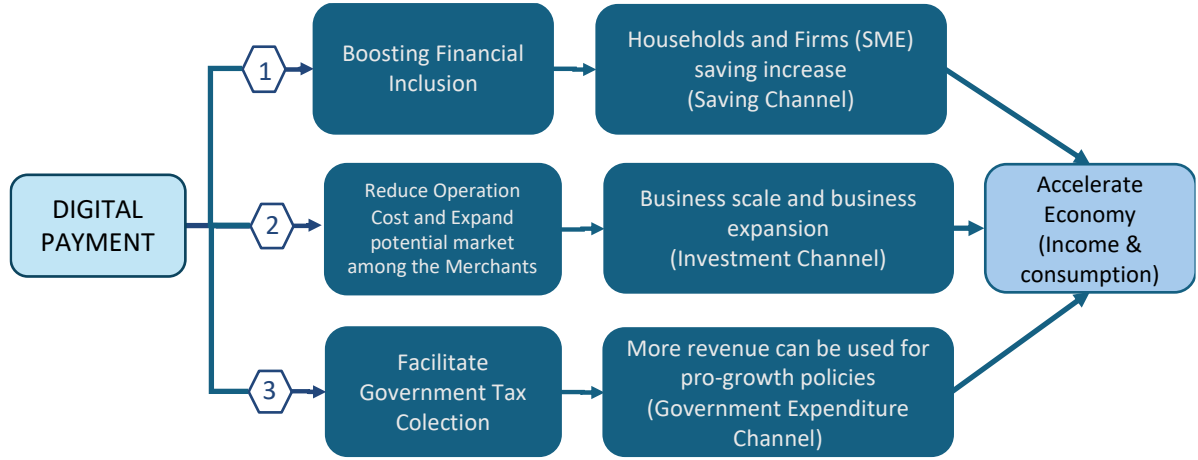

*Source: Author's, developed from Wong et al. (2020)*

Table 1 Descriptive statistics

| Variables | Mean | SD | Min | Max | UR | CD | Obs |
|---|---|---|---|---|---|---|---|
| **Dependen Variable** | | | | | | | |
| GDP | 10.673 | 1.135 | 8.772 | 13.171 | -14.028*** | 115.662*** | 660 |
| Consumption | 10.171 | 1.090 | 8.443 | 12.838 | -15.745*** | 113.913*** | 627 |
| **Independen Variable** | | | | | | | |
| Card Payment | 10.152 | 1.155 | 7.567 | 13.630 | -7.606*** | 60.562*** | 660 |
| ATM/Debit Card | 10.132 | 1.143 | 7.557 | 13.579 | -7.661*** | 60.831*** | 660 |
| Credit Card | 5.641 | 1.933 | 2.128 | 11.016 | -10.307*** | 55.227*** | 660 |
| E-Money | 6.339 | 1.718 | 2.502 | 11.925 | -10.375*** | 81.719*** | 660 |
| RTGS | 9.984 | 1.860 | 4.341 | 16.053 | -9.867*** | 71.477*** | 660 |
| Inflation | 2.709 | 2.117 | -4.82 | 8.620 | -12.692*** | 95.740*** | 660 |
| Merchants | 10.184 | 1.470 | 6.597 | 14.187 | -8.211*** | 65.131*** | 660 |
| Government | 8.313 | 0.833 | 6.539 | 11.293 | -15.930*** | 104.527*** | 627 |
| Capital | 9.508 | 1.136 | 6.840 | 12.188 | -9.632*** | 93.492*** | 627 |
| Human Capital | 71.270 | 4.56 | 20.54 | 82.46 | -14.387*** | 155.645** | 660 |

**Source:** Bank Indonesia. Authors' calculations.
**Note:** Mean, SD, Min, Max, UR, CD, and Obs refer to the sample average, the standard deviation, the minimum value, the maximum value of each variable, unit root test, cross-sectional correlation refers to Pesaran's (2021), and the number of observation, respectively. Panel unit root test was calculated using the **xtbunitroot** syntax as outlined by Chen et al (2021) and Karavias & Tzavalis (2014). The column of variables are logarithm of GDP Real (GDP), logarithm of Consumption (Consumption), logarithm transaction value of all card based payments instrument (Card Payment), logarithm transaction value of ATM and Debit Card (ATM/Debit Card), logarithm transaction value of credit card (Credit card), logarithm transaction value of electronic money (E-money), logarithm transaction value of real time gross settlement (RTGS), Inflasi rate (Inflation), logarithm of number of merchant that accept card and electronic money payments (Merchants), logarithm of government expenditure (Government), logarithm of gross capital formation (Capital), and human development index (Human Capital). Finally, *, ** and *** denote statistical significance at the 10%, 5%, and 1% levels, respectively.

Table 2 Estimated break dates and 95% confidence intervals

| Model and Variabel of Interest | Breaks | GDP | | Consumption | |
|---|---|---|---|---|---|
| | | Date | 95% Conf. Interval | Date | 95% Conf. Interval |
| (1) Card Payment | 1 | 2020q2 | [2020q1, 2020q3] | 2020q2 | [2020q1, 2020q3] |
| (2) ATM/Debit | 1 | 2020q2 | [2020q1, 2020q3] | 2020q2 | [2020q1, 2020q3] |
| (3) Credit Card | 1 | 2020q3 | [2020q2, 2020q4] | 2021q2 | [2021q1, 2021q3] |
| (4) E-Money | 1 | 2020q2 | [2020q1, 2020q3] | 2020q4 | [2020q3, 2021q1] |
| (5) RTGS | 1 | 2022q1 | [2021q4, 2022q2] | 2022q1 | [2021q4, 2022q2] |

**Note:** Break and date refer to the numbers breakpoint and the estimated of breakpoint, respectively. Estimated break dates and 95% confidence interval was calculated using the **xtbreak** syntax as outlined by Ditzen et al. (2021)

Figure 21 Plotting the estimated structural break, digital payments, and regional consumption

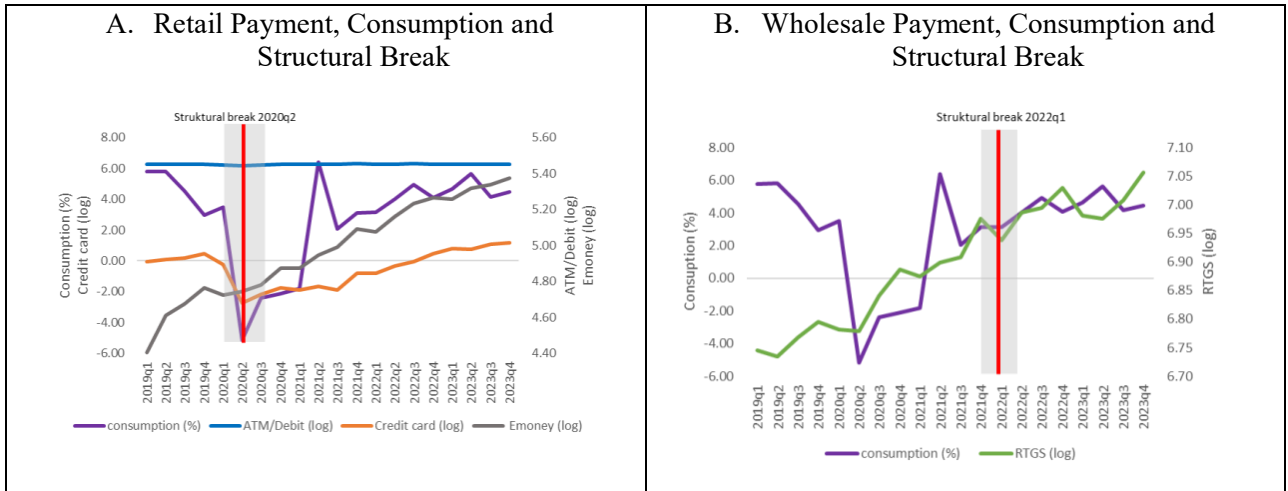

Note: Figure 2 illustrates the evolution of consumption, and digital payment adoption during the observation period, particularly considering the Covid-19 pandemic and the implementation of payment system policy. The structural break date is identified with its 95% confidence interval taken from Table 2.